\documentclass[12pt,preprint]{aastex}

\begin{document}


\title{Indirect ultraviolet photodesorption from CO:N$_2$ binary ices\\
- an efficient grain-gas process}


\author{Mathieu Bertin\altaffilmark{1}, Edith C. Fayolle\altaffilmark{2}, Claire Romanzin\altaffilmark{3}, Hugo A. M. Poderoso\altaffilmark{1}, Xavier Michaut\altaffilmark{1}, Laurent Philippe\altaffilmark{1}, Pascal Jeseck\altaffilmark{1}, Karin I. \"Oberg\altaffilmark{4}, Harold Linnartz\altaffilmark{2} \& Jean-Hugues Fillion\altaffilmark{1}}

\altaffiltext{1}{Laboratoire de Physique Mol\'eculaire pour l'Atmosph\`ere et l'Astrophysique (LPMAA), CNRS UMR 7092, UPMC Univ. Paris 6, F-75252 Paris, France.}
\altaffiltext{2}{Sackler Laboratory for Astrophysics, Leiden Observatory, Leiden University, P.O. Box 9513, NL-2300 RA Leiden, The Netherlands.}
\altaffiltext{3}{Laboratoire de Chimie Physique (LCP), CNRS UMR 8000, Univ. Paris Sud 11, F-91400 Orsay, France.}
\altaffiltext{4}{Harvard-Smithsonian Center for Astrophysics, 60 Garden Street, Cambridge, MA 02138, USA.}

\begin{abstract}
UV ice photodesorption is an important non-thermal desorption pathway in many interstellar environments that has been invoked to explain observations of cold molecules in disks, clouds and cloud cores. Systematic laboratory studies of the photodesorption rates, between 7 and 14 eV, from CO:N$_2$ binary ices, have been performed at the DESIRS vacuum UV beamline of the synchrotron facility SOLEIL. The photodesorption spectral analysis demonstrates that the photodesorption process is indirect, i.e. the desorption is induced by a photon absorption in sub-surface molecular layers, while only surface molecules are actually desorbing. The photodesorption spectra of CO and N$_2$ in binary ices therefore depend on the absorption spectra of the dominant species in the sub-surface ice layer, which implies that the photodesorption efficiency and energy dependence are dramatically different for mixed and layered ices compared to pure ices. In particular, a thin (1-2 ML) N$_2$ ice layer on top of CO will effectively quench CO photodesorption, while enhancing N$_2$ photodesorption by a factors of a few (compared to the pure ices) when the ice is exposed to a typical dark cloud UV field, which may help to explain the different distributions of CO and N$_2$H$^+$ in molecular cloud cores. This indirect photodesorption mechanism may also explain observations of small amounts of complex organics in cold interstellar environments.

\end{abstract}


\keywords{Astrochemistry --- ISM: abundances --- ISM: molecules --- Molecular processes}

\section{Introduction}

In cold and dense regions of the interstellar medium (ISM), characteristic of star and planet formation, gaseous atoms and molecules stick onto dust grains forming ice mantles in relatively short time scales. At these very low temperatures ($<$ 20 K), thermal desorption is negligible for all molecules except H$_2$. Yet, molecules are detected in the gas phase for temperatures below their condensation temperature and this implies the existence of efficient non-thermal desorption processes. These comprise desorption induced by cosmic rays, chemically-induced desorption and vacuum UV photodesorption. The latter has been proposed as an important desorption pathway, particularly in the surface layers of proto-planetary disks \citep{dom05,hog11,wil00}. It may also account for the gas-to-ice abundance ratio for a number of species as observed in other dense regions of the ISM \citep{cou12,hol09}, including the edges of molecular clouds where external UV photons can alter the ice mantle formation, or in the inner part of dense molecular clouds where UV photons are produced by the cosmic-ray ionization of H$_2$ \citep{cas12}. Photodesorption, therefore, is of general importance, influencing the chemistry at different locations in denser regions in space.

Quantitative photodesorption yields from low temperature ices were first obtained experimentally for H$_2$O \citep{wes95a,wes95b}. Over the last five to seven years photodesorption rates have been determined for other pure ice samples using broad band, hydrogen discharge lamps: CO \citep{mun11,obe07}, H$_2$O/D$_2$O \citep{obe09a}, N$_2$ and CO$_2$ \citep{bah12,obe09b,yua13}, O$_2$ and O$_3$ \citep{ZheSub}. Molecular dynamics calculations have been performed for water ice \citep{and05,and06,ara11} and are in good agreement with the experimental findings. More recently, wavelength-specific studies have been performed and for CO, N$_2$ and O$_2$ absolute photodesorption rates have been determined between 7 and 14 eV \citep{fay11,fay13}. The spectral dependence of the photodesorption yield is of fundamental interest for modeling regions in space with different spectral energy distributions. Moreover, this approach has been very successful to link the photodesorption process in the ice to the solid state mechanisms at play, highlighting the physical-chemical parameters governing the photodesorption process. In the case of CO and N$_2$ the wavelength dependent intensity of the measured Photon-Stimulated Desorption signals (PSD) follows directly the electronic transitions in the condensed molecules \citep{fay11,fay13}. This is a signature of a mechanism known as DIET (desorption induced by electronic transition), not (substantially) involving dissociation/recombination in the ices. An even more detailed picture of the involved surface processes has been derived from the investigation of layered $^{13}$CO/$^{12}$CO films \citep{bert12} showing that the photodesorption is mediated by the ice lattice. This study pointed out a sub-surface excitation mechanism in which electronically excited molecules release their energy through intermolecular vibrational motion into a desorption channel. Recently, \cite{yua13} addressed the crucial role of this energy transfer from the ice lattice by comparing Lyman-$\alpha$ photodesorption at 75 K for pure $^{12}$CO$_2$, pure $^{13}$CO$_2$ and $^{12/13}$CO$_2$ mixed ices. 

In the present work we go one step further and study the photodesorption efficiency of a binary ice mixture made of different chemical species. We present wavelength dependent photodesorption measurements from CO and N$_2$ binary ices (layered and mixed) at 10 K. This is interesting from a physical-chemical point of view; their masses are nearly equivalent, their photodesorption mechanisms are similar and their binding energies to the solid are comparable, but the two molecules have strong, non-coinciding photodesorption features in the VUV (around 8.5 and above 12.3 eV for CO and N$_2$, respectively) in their pure solid phase. The choice for a CO:N$_2$ ice is particularly relevant from an astronomical point of view.  CO is highly abundant, both in the gas phase and in the solid state, representing the main molecular component after H$_2$. Molecular nitrogen, N$_2$, is not directly detectable due to its lack of a permanent dipole moment, but is considered as one of the main reservoir of nitrogen in the gas phase, due to its high stability. Its presence in the ISM is evidenced by the detection of N$_2$H$^{+}$ resulting from a proton transfer reaction involving H$_3^{+}$ \citep{berg02,flo06,hil10}. Moreover, N$_2$ stands at the origin of other processes leading to the formation of more complex N-bearing molecules \citep{hil13,per10,per12}. As both CO and N$_2$ are highly volatile, with accretion temperatures in the 16-20 K regime under ISM conditions and sticking coefficients close to unity, they are the last species to freeze onto dust grains, generating a top layer that can be considered as a binary interstellar ice \citep{bis06,obe05}. 

 In this letter we present a laboratory based study of vacuum UV irradiated CO:N$_2$ binary ice samples as a function of the wavelength. This approach follows successful experiments on pure CO, N$_2$ and O$_2$ ice \citep{fay11,fay13,bert12}. In the next section experimental details are given. The results are presented in section 3 and are discussed from a physical-chemical and an astrophysical point of view in section 4.

\section{Experimental methods}
The photodesorption studies are realized in the ‘SPICES’ (Surface Processes \& ICES) setup of the UPMC (Universit\'e Pierre et Marie Curie), under ultrahigh vacuum (UHV) conditions (P $\sim$ 1$\times$10$^{-10}$ Torr). The substrate on which ices are grown is a Highly Oriented Polycrystalline Graphite (HOPG) surface. It is mounted on the tip of a turnable coldhead that can be cooled down to $\sim$ 10 K by means of a closed cycle helium cryostat. The ice layers are grown in-situ by exposing the cold HOPG substrate to a partial pressure of CO and/or N$_2$ gases. A dosing tube, placed 1 mm in front of the substrate, allows a local exposure of the gases onto the cold sample without contaminating the whole UHV chamber. The isotopologues  $^{13}$CO  (Eurisotop, 99.6 \% $^{13}$C) and $^{15}$N$_2$ (Eurisotop, 97 \% $^{15}$N) are used in order for them to be distinguishable by mass spectrometry. The quantities of molecules deposited on the substrate are expressed in ML$_{eq}$ (monolayer equivalent) corresponding to the surface density of a compact molecular layer on a flat surface, with 1 ML$_{eq}$ $\approx$ 1$\times$10$^{15}$ molecule.cm$^{-2}$. Temperature Programmed Desorption (TPD) is used for the calibration of the ice thicknesses, resulting in reproducible parameters for the growth conditions with a precision better than 1 ML$_{eq}$.  

UV photodesorption is induced through irradiation of the ice sample by the continuous output of the undulator-based vacuum UV DESIRS beamline of the synchrotron SOLEIL \citep{nah12}, providing photons with an energy that can be continuously scanned over the 7 - 14 eV range. A narrow bandwidth of typically 40 meV is selected by the 6.65 m normal incidence monochromator that is implemented on the beamline. A gas filter in the beamline suppresses the harmonics of the undulator that can be transmitted in higher diffraction orders of the grating. The absolute incident photon flux per surface unit impinging onto the sample is measured by a calibrated photodiode and varies for a given spectral bandwidth of 40 meV between 0.3 and 1.1$\times$10$^{13}$ photons.s$^{-1}$.cm$^{-2}$ depending on the photon energy. In order to prevent radiation cut-off, the DESIRS beamline is directly connected to SPICES, i.e., without any window. 

The photon-stimulated desorption (PSD) spectra are obtained as follows: the signals of desorbing species with given masses are recorded as a function of time using a Quadrupole Mass Spectrometer (QMS), while the sample is irradiated with photons whose energy is continuously scanned from 7 to 14 eV with steps of 25 meV. This gives the mass signal of desorbing species as a function of the incident photon energy, which can be converted into an absolute photodesorption yield, in desorbed molecule per incident photon, using (\textit{i}) the measured energy-dependent photon flux and (\textit{ii}) the calibration of the mass signal into absolute amount of desorbed molecules. More details on this calibration step have been previously given in \cite{fay11} and \cite{fay13}. It is important to note that this treatment does not significantly change the global shape of the PSD spectra: the observed structures in calibrated PSD spectra are already clearly visible in the unprocessed signals.

Each energy step lasts $\sim$ 5 s. In parallel, the QMS signal is recorded with a dwell of 1 s, which makes an average of 5 points per energy step. Caution has been taken to ensure that for each data point the 1 s accumulation time is substantially longer than the QMS mass signal build-up time, preventing eventual artefacts due to fast wavelength scanning. Each PSD spectra from 7 to 14 eV is made continuously on the same sample. An energy scan from 7 to 14 eV lasts 23 minutes. From our calibrated PSD spectra and the evolution of the photon flux with time and energy, we can estimate the total amount of desorbed molecules per scan for each PSD spectrum. At the maximum, we find a total density of photodesorbed molecules of $\sim$10$^{13}$ molecule.cm$^{-2}$, which corresponds to 10$^{-2}$ ML$_{eq}$. Therefore, the amount of photodesorbed molecules during one single scan is negligible compared to the total amount of molecules available in the ice, which at its lowest value amounts to 0.9 ML$_{eq}$. The photoprocessing of the ice is not expected to modify the photodesorption rate from the beginning to the end of the scan. This is confirmed experimentally, since PSD spectra obtained two times in a row on the same sample are identical.

In order to make a clear separation between signals associated with $^{13}$CO (m = 29 amu) and $^{15}$N$_2$ (m = 30 amu), we verified that our QMS mass resolution is sufficient to differentiate each mass peak. From the mass spectra of the pure products and the mixture of both we derive a resolution of $\sim$ 0.3 amu, as taken from the Full-Width-at-Half-Maximum (FWHM) of the Gaussian mass peaks. It is accurate enough to allow for a fine separation of $^{13}$CO and $^{15}$N$_2$ signals that have indeed a negligible overlap, as can be clearly seen in the corresponding mass spectra, available in appendix A.

\section{Results}

Figure 1a shows PSD spectra obtained upon VUV irradiation of the pure $^{13}$CO and $^{15}$N$_2$ ices. These spectra have already been presented and discussed in previous studies. The photodesorption rates exhibit energy-dependent efficiencies that follow closely the VUV absorption of the pure molecular solids, with patterns that are associated with electronic transitions of the condensed molecules. For instance, the features observed between 7.9 and 9.5 eV in the CO PSD spectrum are due to vibronic bands in the $A^1\Pi - X^1\Sigma^+$(v',0) electronic transition, each peak being associated with a transition towards a vibrational sub-level v' in the A-state of solid $^{13}$CO \citep{fay11}. In a similar way, the main feature responsible for the photodesorption of solid $^{15}$N$_2$ (above 12.3 eV) is associated with the $b^1\Pi_u - X^1\Sigma_g^+$(v',0) transition in pure nitrogen ice \citep{fay13}. The observation of such structures in the energy-resolved photodesorption rates is a clear signature for a DIET mechanism, as discussed in \cite{fay11,fay13}. It has been shown that the excitation mainly takes place in a subsurface region of the ice (2-3 upper molecular layers), while only the molecules from the topmost layer are ejected into the gas phase. A short-range energy transfer from the excited to the desorbing molecules is therefore expected to occur, presumably through the coupling between the relaxation of the excited species and the excitation of intermolecular collective vibrational modes \citep{bert12}. It should be noted that the thickness of our samples is less than the photon penetration depth, implying that photons can excite molecules deeper within the ice and even reach the graphite substrate. However, according to \cite{bert12}, neither the substrate nor underlying layers are contributing to the desorption features observed here. The question that is addressed here is how these properties translate in a mixed ice, consisting of both CO and N$_2$.

\subsection{Photodesorption from mixed CO:N$_2$ ice} 

Figure 1b displays the PSD spectra for both $^{13}$CO and $^{15}$N$_2$ desorbing upon VUV irradiation of an ice grown from a 1:1 CO:N$_2$ mixture. There is a clear difference between the results for the pure and mixed ice irradiation. As discussed above, the pure ice spectra are very different, whereas, for the mixed ice, the PSD spectra are identical. A comparison with Fig. 1a shows that the photodesorption spectra of the CO:N$_2$ ice mixture results in a superposition of the PSD spectra of the two pure constituents. The ejection of any surface molecule in the mixed ice, clearly, is initiated by the electronic excitation of any other molecule. In particular, the PSD spectra of the mixed ice show that the excitation of solid CO into its $A^1\Pi$ state leads to the desorption of surface N$_2$ (7.9 - 9.5 eV), and that solid N$_2$ excitation into its $b^1\Pi_u$ state initiates CO desorption above 12.3 eV. This finding is fully consistent with the previously introduced concept of an indirect DIET mechanism; part of the excess energy deposited in the ice by the vacuum UV photon absorption is transferred from one to the other molecule – independently of its chemical nature - causing desorption.  

For a given energy, the absolute photodesorption rates of CO and N$_2$ from the mixed ice are lower by a factor of 2 than the corresponding value derived for the pure ices. In fact, the PSD spectra of the mixture can be fitted with a very good agreement by a linear combination of the desorption spectra of pure $^{15}$N$_2$ and pure $^{13}$CO in which each constituent contributes to about $\sim$ 0.5 of the overall desorption signal (Fig. 2). As the surface of the mixed film is expected to be composed of half CO and half N$_2$ molecules, and as the adsorption energies of CO and N$_2$ are very close \citep{bis06}, this shows that the amount and efficiency of the energy transfer to surface molecules upon subsurface CO excitation or N$_2$ excitation are about the same.

\subsection{Photodesorption from layered N$_2$/CO and CO/N$_2$ ices} 

Photon-stimulated desorption spectra of $^{13}$CO (left panel) and $^{15}$N$_2$ (right panel) obtained from a pure 25 ML$_{eq}$ $^{13}$CO ice covered by an increasing layer of $^{15}$N$_2$ are shown in Fig. 3. In the case of pure CO or N$_2$ ice (top left and bottom right, respectively) the characteristic desorption profiles around 8.5 and above 12.3 eV are seen. When the CO ice is covered by 0.9 ML$_{eq}$ of N$_2$, the CO desorption yield drops substantially, and it is almost entirely suppressed above ML$_{eq}$. This behavior, also observed in the case of layered $^{13}$CO/$^{12}$CO \citep{bert12}, shows that mostly the surface molecules are susceptible to desorb upon UV irradiation. 

When investigating the PSD of $^{15}$N$_2$ for a low thickness of N$_2$ overlayer (0.9 ML$_{eq}$), one can see that its photodesorption pattern mimicks almost completely the one of the pure CO ice, demonstrating that its photodesorption is triggered only by the absorption of the underlying CO molecules. In particular, its photodesorption in the 7.9 - 9.5 eV range, associated with the excitation of the $A^1\Pi$ state of solid CO, becomes the dominant contribution, although pure solid N$_2$ does not photodesorb at these energies. In contrast, photodesorption initiated by the N$_2$ excitation in the $b^1\Pi_u$ state ($>$ 12.3 eV) is not clearly observed. This cannot be explained by the lack of surface N$_{2}$ at the end of the energy scan, since less than 0.01 ML$_{eq}$ of N$_{2}$ has been ejected into the gas phase during the acquisition time. The adsorption of a small quantity of N$_2$ onto a CO ice therefore drastically modifies the energy-dependence of its photodesorption process. For increasing thickness of the $^{15}$N$_2$ overlayer, the N$_2$ excitation contribution to the N$_2$ PSD spectra gradually recovers the one of the pure N$_2$ ice. The contribution of the CO excitation becomes very weak for N$_2$ thicknesses above ML$_{eq}$, showing that the range for an efficient energy transfer involves less than 3 molecular layers. 

Figure 4 shows the results of layered experiments in which 25 ML$_{eq}$ $^{15}$N$_2$ ice is covered by an increasing layer of $^{13}$CO. The experiment is essentially identical to the previous one (fig. 3), but the role of N$_2$ and CO are exchanged. The N$_2$ photodesorption signal vanishes for a larger CO coverage and a clear vibrational progression in the $^{13}$CO desorption channel is found that corresponds to an excitation into the $b^1\Pi_u$ state of N$_2$. This feature is superimposed over a continuous desorption of $^{13}$CO observed above 12.5 eV in pure CO samples (Fayolle, et al. 2011). As for the 25 ML$_{eq}$ $^{13}$CO ice covered by $^{15}$N$_2$ the spectra are fully consistent with a process in which only surface molecules desorb after excitation of sub-surface species.

\section{Discussions and astrophysical implications}

\subsection{Role of ice structure and composition on absolute photodesorption rates} 

The VUV photodesorption of coadsorbed N$_2$ and CO presents very different energy-dependent profiles and efficiencies compared to pure N$_2$ and CO ices. Both molecules mainly desorb through an indirect mechanism, where the excitation of a subsurface molecule, whether CO or N$_2$, leads to the desorption of N$_2$ and CO surface molecules. Because this indirect mechanism involves energy transfer from the excited molecule to the desorbing one, presumably by collective vibrational mode excitation, the photodesorption energy-dependence and efficiency depend strongly on the nature of the intermolecular interactions. In general intermolecular interactions in solids may promote the photodesorption, but also quench it. This has been demonstrated for CO$_2$ in interaction with other isotopologues or rare gas matrices \citep{yua13}, and for CO in interaction with H$_2$O ice \citep{bert12}. For the latter case, an explanation has been proposed, stating that the photodesorption efficiency is driven by the competition between two excess energy relaxation pathways: (\textit{i}) relaxation by intermolecular phonon modes of the ice, coupled to desorption and (\textit{ii}) relaxation by transfer to intramolecular vibration modes of neighboring molecules. If the first relaxation pathway is dominant, the photodesorption will be efficiently triggered by UV absorption of neighboring molecule. If the latter pathway dominates the relaxation process, as it is the case when excited CO transfers efficiently its excess energy to the O-H dangling bond vibrations of neighboring water molecules, then the photodesorption gets hindered. In both cases, the constructive and destructive matrix effects are a consequence of the indirect character of the photodesorption mechanism. 

Because the intermolecular coupling in the ice between excited and surface molecules can drastically modify the photodesorption process, it is expected that different composition, but also different structure of the ice, will modify the energy-integrated photodesorption yield for a given UV field. This is illustrated for the N$_2$/CO binary ices. As shown in Figs. 1 and 3, coadsorption of N$_2$ with CO promotes N$_2$ desorption in the 7 - 11.5 eV range whereas it decreases the desorption efficiency of CO in the same energy range. To quantify this effect, energy-integrated photodesorption rates have been derived from our PSD curves for three types of interstellar vacuum UV profiles as a function of the CO:N$_2$ layered/mixed ices composition. The results are displayed in Table 1. To obtain these values, our energy-resolved desorption rates have been convoluted with vacuum UV profiles describing the interstellar radiation field (ISRF) at the edge of clouds \citep{mat83}, the effect of secondary photons from cosmic ray impacts in dense cores \citep{gre87}, and a TW Hydr\ae~radiation field as model for the UV field in proto-planetary disks \citep{herc02,val03,joh07}. The resulting photodesorption rates are varying depending on (\textit{i}) the UV field considered, (\textit{ii}) the ice composition and (\textit{iii}) the internal ice organization (layered versus mixed). Compared to pure ices the photodesorption rates from the binary ices change by up to an order of magnitude. The general trend is to decrease the CO photodesorption rate from $>$ 1 $\times 10^{-2}$ to 2-6 $\times 10^{-3}$ molecules.photon$^{-1}$, and to increase the N$_2$ photodesorption rate by a factor of 2-3. This shows that a proper description of the VUV photodesorption process has to take into consideration the structure and composition of the ice on which the molecules are adsorbed; the photodesorption rates extracted from pure molecular solids may not be realistic in the case of more complex ices, as typically present in the insterstellar medium.

\subsection{Implications for N$_2$ and CO gas phase abundance in dense cores} 

In dense cores, N$_2$ and CO are expected to freeze out onto dust grains at very low temperatures, less than 2 K apart, and with the same sticking efficiency (Bisschop et al. 2006; Oberg et al. 2007). Consequently, a similar depletion in the gas phase should be observed. Nevertheless, observational studies of N$_2$H$^+$ and other N-containing species in dense cores providing data to indirectly derive the N$_2$ abundance suggest that  N$_2$ depletion on grains occurs later, i.e. at a higher density, than for CO \citep{berg02,pag05,pag12}. Therefore, it has been suggested that another mechanism is repsonsible for a continuous enrichment of gas phase N$_2$, counterbalancing its accretion onto grains. Chemistry involving CN + N reactions, constantly forming N$_2$ in the gas, is invoked \citep{hil10}. Non-thermal desorption processes, such as photodesorption, could also contribute to this effect by preferentially ejecting N$_2$ instead of CO in the gas phase. Nevertheless, the photodesorption rates as extracted from pure CO and N$_2$ ices cannot resolve this issue since for realistic dark cloud UV field the photodesorption rate is substantially lower for pure N$_2$ than for pure CO ice (Table 1). However, considering pure ices only may be inappropriate, as discussed in section 4.1, since ice structure and composition influence the CO and N$_2$ photodesorption rates.

 Since CO and N$_2$ are expected to freeze-out in the same temperature range, it is rather unlikely that the top layers of the ices are composed solely of pure N$_2$ or pure CO thick layers. Instead, we expect that both species are coadsorbed at the surface of the icy grains. The exact nature of the top layers depends on the total abundance of N$_2$ as compared to CO, as well as on the proximity in time and space of CO and N$_2$ freeze-out. The amount of gas phase N$_2$ being relatively low compared to CO \citep{berg02,mar06}, it is expected that solid phase N$_2$ will be embedded in a CO-rich environment. Whether the outer layers of the grain will be layered or mixed is not that easy to answer, and is likely source-dependent. N$_2$ is expected to condense for a temparture that is 2 K lower than for CO \citep{flo06,hil10}, which would, for low cooling speed of the cloud, lead to the growth of a thin N$_2$ layer on top of CO-rich ice. In the case of a shorter cooling time, such a temperature difference would not be sufficient to result in segregated layers, and N$_2$ would be intimately mixed with the CO.

 Table 1 shows that for realistic dark cloud UV field photodesorption rates extracted from pure N$_2$ ice is substantially lower for than for pure CO ice. If we consider instead mixed CO:N$_2$ 1:1 ice, the photodesorption rates of CO and N$_2$ in prestellar cores are identical. In the case of a more realistic system, i.e. a small amount (0.9 ML$_{eq}$) of N$_2$ on CO ice, then, the CO photodesorption rate strongly decreases, whereas the N$_2$ photodesorption rate is promoted. Then, the N$_2$ desorption becomes more efficient than the CO one, and photodesorption can contribute to gas phase N$_2$ enrichment. We conclude that under interstellar conditions N$_2$ will be photodesorbed at least as efficiently as CO as long as they are coadsorbed. In many core environments, the N$_2$ photodesorption may exceed that of CO providing a natural explanation to why N$_2$ is maintained longer in the gas phase compared to CO.

\subsection{Organics desorption through CO photoexcitation?} 

The indirect nature of the photodesorption process may have general implications beyond the explanation of the CO and N$_2$ abundances in the ISM. The ability of CO ice to transfer part of the absorbed energy to a surrounding molecule may act as an important indirect photodesorption channel of other species as well. This could be true in particular for chemically linked formaldehyde H$_2$CO and methanol CH$_3$OH, and possibly even more complex organic molecules, providing a non-dissociative non-thermal desorption pathway. Indeed, such organics are expected to efficiently photodissociate upon UV photon irradiation, as it has been found for CH$_3$OH \citep{obe09c}, leading to the formation of photoproducts that can eventually desorb.  Embedded in a CO-rich environment, however, such molecules may react differently under UV irradiation: a photon absorption by the surrounding and more numerous CO molecules could lead to the desorption of the surface-located organics via a similar mechanism as highlighted for CO:N$_2$. Such a process could then explain the observed gas abundances of organics, observed in the cold regions of the ISM \citep{bac12,obe10,ter13}. Whether this indirect photodesorption process can be generalized to heavier and more strongly bound species than CO and N$_2$ needs to be experimentally verified, however.

\section{Summary and conclusions}

		From a systematic set of energy (7 - 14 eV) resolved (40 meV) photodesorption laboratory studies on CO:N$_2$ binary ices, both mixed and layered the following conclusions can be made:
\begin{itemize}
\item The desorption is induced by a photon absorption in the topmost molecular layers, while only surface molecules are actually desorbing. This implies an energy transfer from the subsurface excited molecule to the surface ones. Thus, the photodesorption rate is not linked to the absorption spectrum of the desorbing molecule, but is instead associated to the absorption profile of the surrounding species, located deeper in the ice. When N$_2$ and CO are mixed in equal proportions, their photodesorption spectra become superimposable, and reflect a linear combination of the photodesorption spectra of both pure ices. When a small quantity of N$_2$ is deposited at the surface of a CO ice, then, its photodesorption spectrum reflects mostly the one of pure CO. This is remarkable since, in this case, the N$_2$ major desorption feature lies in the 7.9 - 9.5 eV energy range, in which pure solid N$_2$ does not absorb. 

\item This indirect desorption mechanism and its dependence on ice structure and wavelength influence the (overall) desorption efficiency in space. Considering the range of indirect photodesorption processes, we suggest that the photodesorption rates experimentally obtained from pure ices should only be used for molecules which can form pure phases in the ISM ices, exceeding at least two to three molecular toplayers, as it may be the case for CO in some cold regions \citep{pon08,pon03}. Below this thickness, or when molecules are mixed with others, photodesorption rates obtained from a more realistic composite ice should lead to a more accurate description of the non-thermal desorption. Here both chemical composition and molecular organization play a role.

\item The results presented here provide a way to replenish nitrogen in the prestellar cores, contributing to the unexplained low depletion rate of N$_2$ as compared to CO. If the rates derived for pure CO and N$_2$ are used, CO should desorb with an efficiency that is almost one order of magnitude higher than for N$_2$. This is not in agreement with observations concluding that gaseous CO depletes on the grains with a higher rate than N$_2$ \citep{pag12}. When considering the rates extracted from a more realistic system, most probably a layered N$_2$/CO structure with a small amount of N$_2$, as presented here, gaseous N$_2$ will be enriched with respect to CO. A detailed astrochemical modelling is required to quantify the extent of this N$_2$ enrichment.

\item Finally, the mechanism introduced here may act as a more general desorption process in other mixed ices of astrophysical interest. The CO present in CO-rich icy mantles has the potential to transfer photon energy to kinetic energy of surrounding molecules, triggering their photodesorption in an indirect way. The concerned species could be co-adsorbed molecules during the freeze-out process, such as unsaturated carbon chains, but also small organics originating from the hydrogenation of CO, such as H$_2$CO and CH$_3$OH.
\end{itemize}

\acknowledgments

We acknowledge SOLEIL for provision of synchrotron radiation facilities under the project 20120834 and we would like to thank Nelson De Oliveira for assistance on the beamline DESIRS. Financial support from the French CNRS national program PCMI (Physique et Chimie du Milieu Interstellaire), the UPMC platform for astrophysics "ASTROLAB", the Dutch program NOVA (Nederlandse Onderzoekschool voor Astronomie), and the Hubert Curien Partnership Van Gogh (25055YK) are gratefully acknowledged. The Leiden group acknowledges NWO support through a VICI grant.

\clearpage

\begin{figure}

\plotone{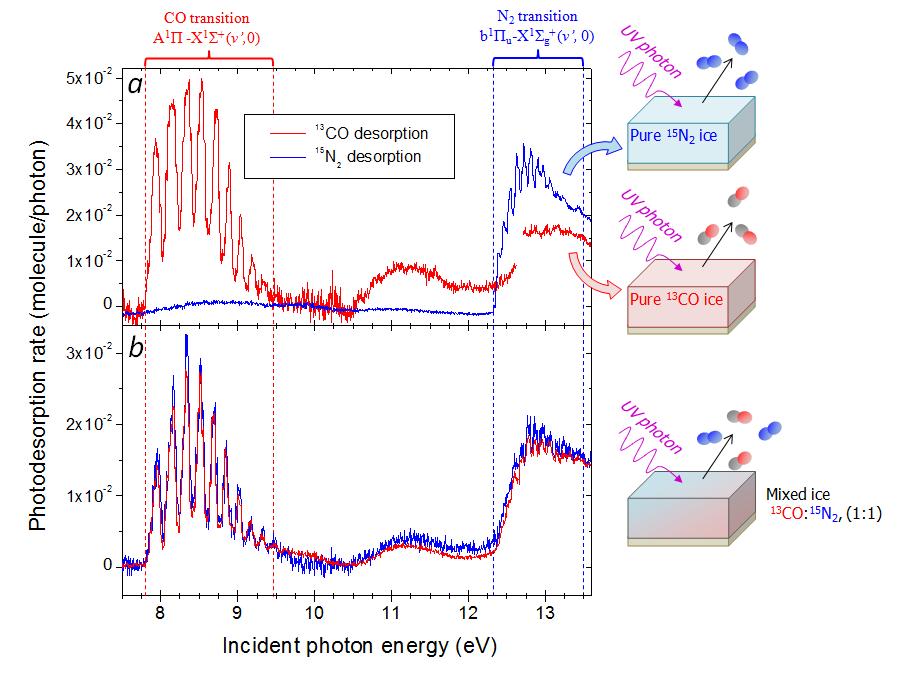}
\caption{Photon-stimulated desorption spectra of $^{13}$CO (red) and $^{15}$N$_2$ (blue) from 30 ML$_{eq}$ of pure $^{13}$CO and $^{15}$N$_2$ ices (a), and from a mixed $^{13}$CO:$^{15}$N$_2$ ice, in proportion 1:1 (b). The electronic transitions in condensed CO and N$_2$ associated with the main photodesorption features are indicated. All spectra have been obtained for ices kept at 15 K and deposited on HOPG. }.\label{fig1}
\end{figure}

\clearpage

\begin{figure}

\plotone{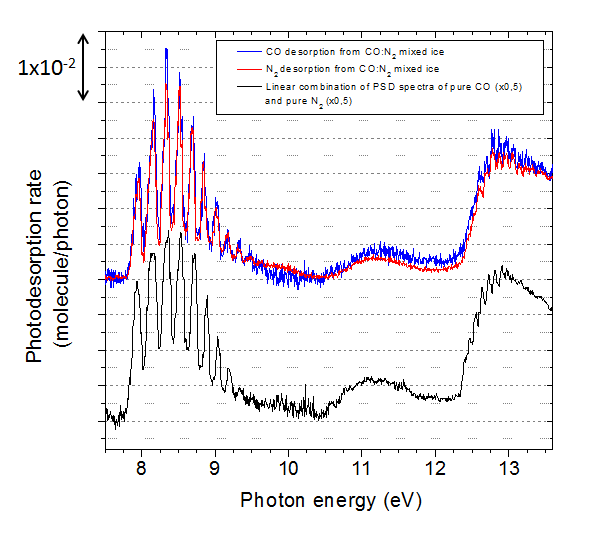}
\caption{Photon-stimulated spectra of $^{15}$N$_2$ and $^{13}$CO from a mixed $^{13}$CO:$^{15}$N$_2$ 30 ML$_{eq}$ ice, in proportion 1:1, compared with a linear combination of PSD spectra obtained from pure $^{13}$CO ice ($\times$ 0.5) and pure $^{15}$N$_2$ ice ($\times$ 0.5). The spectra have been acquired from ices deposited on HOPG kept at 15 K.}.\label{fig2}
\end{figure}

\clearpage

\begin{figure}

\plotone{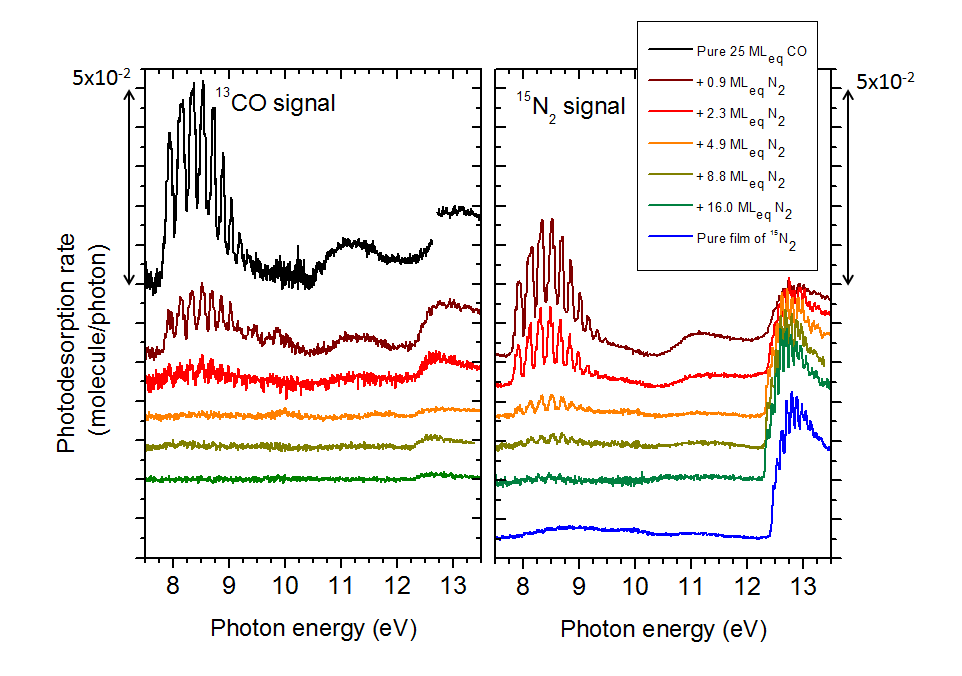}
\caption{Photon-stimulated desorption spectra of $^{13}$CO (left) and $^{15}$N$_2$ (right) from 25 ML$_{eq}$ of $^{13}$CO ice covered by an increasing $^{15}$N$_2$ layer on top. The PSD spectrum obtained from a pure 25 ML$_{eq}$ $^{15}$N$_2$ ice is also presented for comparison. All spectra are recorded for 15 K ices, deposited on HOPG.}.\label{fig3}
\end{figure}

\clearpage

\begin{figure}

\plotone{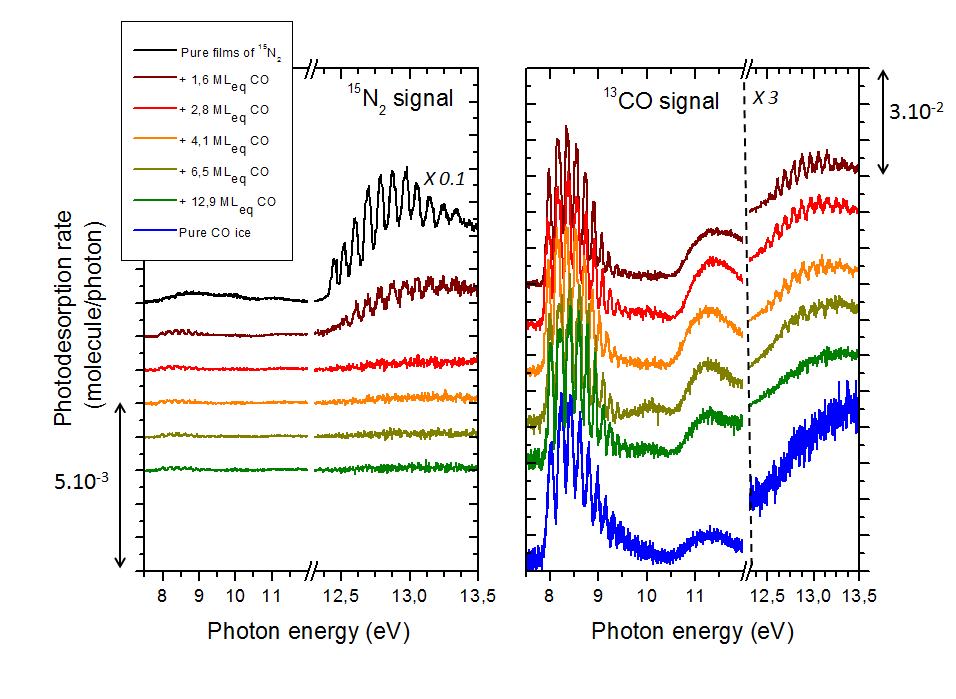}
\caption{Photon-stimulated desorption spectra of $^{15}$N$_2$ (left) and $^{13}$CO (right) from 25 ML$_{eq}$ of $^{15}$N$_2$ ice covered by an increasing layer of $^{13}$CO ice on top. All spectra are recorded for 15 K ices, deposited on HOPG. The top spectra of the left panel (photodesorption of N$_2$ from a pure N$_2$ ice) has been multiplied by 0.1 to facilitate the reading. In the right panel, the CO PSD spectra in the 12.3 - 13.5 eV energy range have been multiplied by 3.}\label{fig4}
\end{figure}

\clearpage

\begin{table}
\begin{center}
\caption{Energy-integrated photodesorption rates of CO and N$_2$ in several regions of the ISM for pure ices, mixed CO:N$_2$ ices in a proportion of 1:1, and 0.9 monolayer equivalent (ML$_{eq}$) of N$_2$ deposited on top of a CO ice. The rates are for ices kept at 15 K. All values are given in desorbed molecules per incident photon. \label{tab1} }
\begin{tabular}{lcccccc}     
\tableline
\tableline       
 & Pure  & Pure & \multicolumn{2}{c}{CO from} &  \multicolumn{2}{c}{N$_2$ from} \\ Environment &  CO ice$^d$ & N$_2$ ice$^e$ & Mixture & 0.9 ML$_{eq}$ & Mixture & 0.9 ML$_{eq}$ \\ &  &  & CO:N$_2$ 1:1 & N$_2$ on CO & CO:N$_2$ 1:1 & N$_2$ on CO \\
\tableline
Edges of clouds$^a$ & $1.3 \times 10^{-2}$ & $2.6 \times 10^{-3}$ & $5.7 \times 10^{-3}$ & $5.3 \times 10^{-3}$ & $5.5 \times 10^{-3}$ & $8.0 \times 10^{-3}$ \\
Prestellar cores$^b$ & $1.0 \times 10^{-2}$  & $2.2 \times 10^{-3}$ & $3.0 \times 10^{-3}$ & $3.9 \times 10^{-3}$ & $3.0 \times 10^{-3}$ & $5.1 \times 10^{-3}$ \\
Protoplanetary disk$^c$ & $7.2 \times 10^{-2}$  & $5.3 \times 10^{-3}$ & $2.3 \times 10^{-3}$ & $3.0 \times 10^{-3}$ & $2.1 \times 10^{-3}$ & $2.7 \times 10^{-3}$ \\
\tableline
\end{tabular}
\tablecomments{Using UV field from $^a$Mathis et al. 1983, $^b$Gredel et al. 1987, $^c$Herczeg et al. 2002, Valenti et al. 2003, Johns-Skrull \& Herczeg 2007. $^d$ and $^e$ are values from Fayolle et al. 2011 and Fayolle et al. 2013, respectively.}
\end{center}
\end{table}

\clearpage

\appendix
\section{Appendix: Separation by mass spectrometry of $^{15}$N$_2$ and $^{13}$CO signals}

	The study of the photodesorption of N$_2$ and CO from mixed ice requires a fine separation of the gas phase signals associated with each molecule. In their most abundant isotopic form, $^{14}$N$_2$ and $^{12}$C$^{16}$O, both species have the same mass, and cannot be separated by mass spectrometry. We therefore have used the isotopologues $^{13}$C$^{16}$O and $^{15}$N$_2$ with respective mass 29 uma and 30 uma. To ensure that the rise of one given mass signal does not lead to the artificial increase of the neighbouring mass channel, it is necessary to check that the width of each mass signal is narrow enough to prevent significant overlap between the associated mass peaks.
	Figure A shows mass spectra, i.e. signal of the singly ionized gas phase species as a function of its mass, realized when a pressure of (\textit{i}) pure $^{15}$N$_2$, (\textit{ii}) pure $^{13}$CO, and (\textit{iii}) mixed $^{13}$CO:$^{15}$N$_2$ in proportion 1:1, is introduced into the chamber. The mass peaks of $^{15}$N$_2$ (m = 30 uma) and $^{13}$CO (m = 29 uma) are clearly separated. The rise of one mass signal does not lead to a measurable increase of the other mass channel. Moreover, the mass spectrum associated with the gaseous mixture exhibits no overlap between the two mass peaks. In the photodesorption experiments that are presented, the signals on the mass channel 29 uma or 30 uma can therefore be associated without ambiguity to the presence into the gas of $^{13}$CO or $^{15}$N$_2$, respectively.

\begin{figure}
\figurenum{A}
\plotone{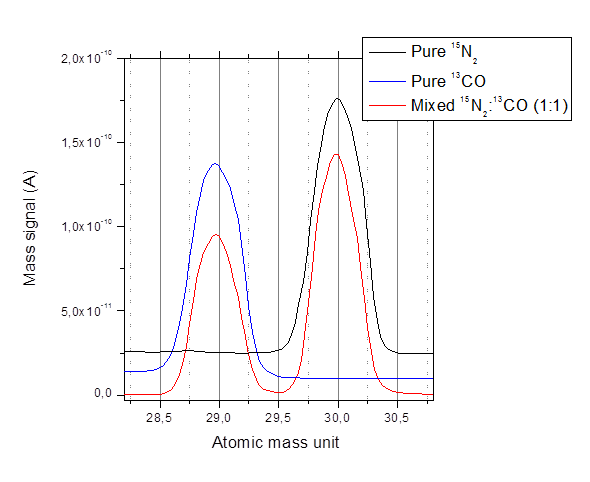}
\caption{Mass spectra obtained from introduced partial pressure of (\textit{i}) $\sim$ 10$^{-9}$ Torr of $^{15}$N$_2$, (\textit{ii}) $\sim$ 10$^{-9}$ Torr of $^{13}$CO, and (\textit{iii}) a mixture of $\sim$ 10$^{-9}$ Torr of $^{15}$N$_2$ + $\sim$ 10$^{-9}$ Torr of $^{13}$CO.}.\label{figA}
\end{figure}

\end{document}